\documentclass[superscriptaddress,reprint, aps, pre]{revtex4-1}%leqn, 
\usepackage[english]{babel}
\usepackage{graphicx}
\usepackage{amsfonts}
\usepackage{amssymb}
\usepackage{amsmath}
\usepackage{hyperref}
\usepackage{cancel}
\usepackage{colortbl}
\usepackage{siunitx}
\usepackage{soul}				%underline
\usepackage{picture}
\usepackage{rotating}
\usepackage{textcomp}
\usepackage{transparent}

\usepackage{graphicx}
\usepackage{epstopdf}
\usepackage[abs]{overpic}
\usepackage{xcolor,varwidth}
\DeclareSymbolFontAlphabet{\mathbb}{AMSb}
\newcommand{\eqr}[1]{Eq.~\eqref{#1}}

\newcommand{\kbt}{k_{B}T}

\newcommand{\rhos}{\rho_{\mathrm{s}}}
\newcommand{\lb}{\lambda_{B}}
\newcommand{\ld}{\lambda_{D}}
\newcommand{\ls}{\lambda_{S}}

\newcommand{\nn}{\nonumber\\}

\begin{document}

\author{Mathijs Janssen}
\email{mjanssen@is.mpg.de}
\author{Markus Bier}
\email{bier@is.mpg.de}
\affiliation{Max Planck Institut f\"{u}r Intelligente Systeme, Heisenbergstr. 3, 70569 Stuttgart, Germany}
\affiliation{Institut f\"{u}r Theoretische Physik IV, Universit\"{a}t Stuttgart, Pfaffenwaldring 57, 70569 Stuttgart, Germany}

\date{\today}

\begin{abstract}
We revisit a classical problem of theoretical electrochemistry: 
the response of an electric double layer capacitor (EDLC) subject to a small, suddenly applied external potential.
We solve the Debye-Falkenhagen equation to obtain exact expressions for key EDLC quantities: 
the ionic charge density, the ionic current density, and the electric field. In contrast to earlier works, our results are not restricted to 
 the long-time asymptotics of those quantities.  
The solutions take the form of infinite sums whose successive terms all decay exponentially 
with increasingly short relaxation times.
Importantly, this set of relaxation times is the same among all aforementioned EDLC quantities; this property is demanded on
physical grounds but not generally achieved within approximation schemes.  
The scaling of the largest relaxation timescale $\tau_{1}$, that determines the long-time decay, is in accordance with earlier results: 
Depending on the Debye length, $\ld$, and the electrode separation, $2L$,
it amounts to $\tau_{1}\simeq\ld L/D$ for $L\gg\ld$, and $\tau_{1} \simeq 4 L^2/(\pi^2 D)$ for $L\ll\ld$, 
respectively (with $D$ being the ionic diffusivity). 
\end{abstract}

\title{Transient dynamics of electric double-layer capacitors: Exact expressions within the Debye-Falkenhagen approximation}
\maketitle

\section{Introduction}
\begin{figure}[b]
\centering
\includegraphics[width=0.48\textwidth]{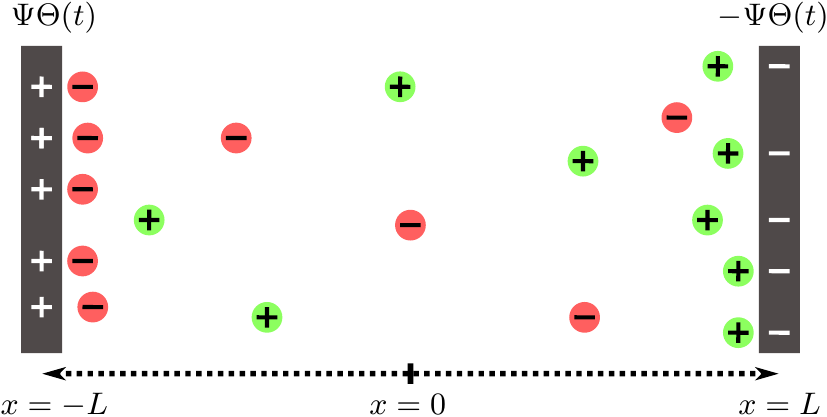}
\caption{A model EDLC consisting of a 1:1 electrolyte and two flat electrodes separated over a distance $2L$. Here $\Theta(t)$ is the Heaviside 
function that takes the time $t$ as an argument; at $t=0$, a potential difference $2\Psi$ is applied.}
\label{fig1}
\end{figure}

Understanding the time-dependent formation of electric double layers (EDLs) in response to varying external influences is a 
fundamental problem of relevance to diverse fields including electrochemistry \cite{bagotsky2005fundamentals, schmickler2010interfacial}, colloid science \cite{russel1989colloidal, hunter2001foundations}, 
biophysics \cite{jue2017modern, ashrafuzzaman2012membrane}, 
and microfluidics \cite{RevModPhys.77.977}. 
Moreover, the speed with which EDLs can form in so-called electric double-layer capacitors (EDLCs) crucially determines the feasibility of these 
devices for energy storage \cite{simon2008materials} and conversion of energy \cite{brogioli2009extracting, janssen2016harvesting}.
The starting point in any classical treatment of dynamics in ionic fluids are the Poisson-Nernst-Planck (PNP) equations
--a set of coupled differential equations that capture the time-varying electric potential and ionic densities \cite{planck1890ueber}.
Then, the canonical model setup (see Fig.~\ref{fig1}) for studying ionic dynamics is that of an electrolyte 
confined by two parallel flat electrodes separated over a distance $2L$.
With this setup, people have studied the formation of EDLs in reaction to a sudden change in chemical 
environment \cite{cabaleiro2014transient} and in the temperature at the electrodes \cite{stout2017diffuse}.
But the canonical problem, especially after the seminal paper of Bazant \textit{et al.} \cite{bazant2004diffuse}, 
is that of an electrolyte subject to a suddenly applied potential difference between the electrodes (Ref.~\cite{bazant2004diffuse} also contains an exhaustive
historical review on prior work on diffuse charge dynamics). 
Later work has considered electrode porosity \cite{biesheuvel2010nonlinear},
heat production caused by finite ionic currents \cite{janssen2017reversible},
and adsorption \cite{alexe2006transient, barbero2006role} and 
Faradaic reactions \cite{barbero2007electrical, van2010diffuse} at the electrode surfaces. 
Moreover, with various analytical and numerical techniques, people have studied the 
PNP equations at large applied potentials \cite{bazant2004diffuse, alexe2006transient, barbero2006role, beunis2008dynamics, olesen2010strongly, morrow2006time,lim2007transient,hossain2013dynamic},
 giving rise, i.a., to neutral salt diffusion, which is of special interest to many practical situations and applications.
However, in modern supercapa\-ci\-tive devices,
nanoporous carbon electrodes are charged to such high potentials (up to $\approx2.5$~V)
that steric repulsions between the ions and the electrodes and between the ions themselves start
affecting the local ionic densities, 
leading for instance to ionic layering perpendicular to the electrode surfaces \cite{hartel2017structure}. 
Since such effects cannot be captured with the original PNP equations, 
later work developed various modifications to these equations \cite{kilic2007stericII, kondrat2015dynamics}, or resorted to
dynamical density functional theory \cite{jiang2014time} or simulations  \cite{cagle2010structure, thakore2015charge} 
to describe ionic relaxation under strong confinement and at high potentials.

Notwithstanding these efforts to develop ever more accurate descriptions of ionic relaxation 
in situation relevant to practical applications and devices, 
the present manuscript concerns with the first model problem posed in Ref.~\cite{bazant2004diffuse}:
the model EDLC of Fig.~\ref{fig1} subject to a suddenly applied potential smaller than the thermal voltage. 
Under these conditions, the PNP equations give rise to the Debye-Falkenhagen (DF) equation [cf. \eqr{eq:DebyeFalkenhagen}]: 
a drift-diffusion equation for ionic charge density \cite{debye1928dispersion}.
This equation has been solved under various assumptions and ansatzes \cite{bazant2004diffuse, alexe2006transient, barbero2006role}.
Specifically, Ref.~\cite{bazant2004diffuse} applied a Laplace transform on the time variable of the DF equation, 
which transforms the PDE for the local ionic charge density into an ODE [cf.~\eqr{eq:laplacechargedensity0}], which is easily solvable.
However, the inverse Laplace transform, required to find the real-time ionic charge density, is notoriously difficult. 
Reference~\cite{bazant2004diffuse} proceeded by applying a so-called Pad\'{e} approximation, 
essentially molding the Laplace-transformed ionic charge density into a form whose inverse Laplace transform is tabulated.  

Later works have proposed 
solutions to the DF equation \cite{golovnev2009exact, golovnev2011analytical}, as well as 
solutions for limiting cases of high and low salt content \cite{beunis2008dynamics}. 
However, all these works have circumvented directly performing the inverse Laplace transform on the ionic charge density
because of its perceived analytical difficulty \cite{bazant2004diffuse} or asserted impossibility \cite{golovnev2009exact}.
In this article we show that this inverse Laplace transformations is, in fact, possible; 
we report new expressions for the ionic charge density, the ionic current density, 
and the electric field of a model EDLC subject to a small, suddenly applied potential difference. 
Our expressions take the form of infinite sums with coefficients depending on the solutions $\mathcal{M}_{j}$ of a transcendental equation. 
These $\mathcal{M}_{j}$ simplify, however, for the limiting 
case of strong double layer overlap ($\ld\gg L$, with $\ld$ being the salt concentration-dependent Debye length), which is relevant, e.g.,
to nonpolar solvents that allow very low salt concentrations.
In that case, our expression for the ionic charge density 
reproduces the exact solution implicit in Ref.~\cite{beunis2008dynamics}. 
But our expressions work equally well away from this limiting case: 
They are in excellent agreement with numerical inverse Laplace transformations 
for all times and system sizes considered. 
Importantly, the aforementioned time-dependent EDLC properties 
all decay with the same set of relaxation timescales $\tau_{j}$. 
This property, not satisfied   within the aforementioned Pad\'{e} approximation scheme,
is physically demanded on the basis of the Poisson and continuity equations.
We confirm previously found scaling of the long-time relaxation timescale $\tau_{1}$ for thin ($\ld\ll L$) 
and thick ($\ld\gg L$) double layers, which read $\tau_{1} \simeq \ld L/D$ and $\tau_{1}\simeq 4L^2/(\pi^2 D)$, respectively.

This article is structured as follows. We describe the setup and governing equations in Sec.~\ref{sec:setup}.
Section~\ref{sec_pade}  reviews the Pad\'{e} approximation scheme employed by previous authors, and highlights its problematic implications.
In Sec.~\ref{sec:inverse}, we perform inverse Laplace transformations to obtain exact expressions for the 
ionic charge density, ionic current density, and electric field, which are discussed and compared to earlier results in Sec.~\ref{sec:discussion}. 
Besides concluding remarks, Sec.~\ref{sec:conclusion} contains suggestions for future work.

\section{Setup}\label{sec:setup}
We consider a cell (see Fig.~\ref{fig1}) consisting of a dilute 1:1 electrolyte solution at a constant temperature 
$T$ bound by two flat, blocking electrodes at $x=-L$ and $x=L$, with $L$ much larger than the size of the electrolyte molecules. 
We treat the solvent as a homogenous dielectric background of constant relative permittivity $\varepsilon_{r}$, 
thus ignoring the possibly intricate dependence of $\varepsilon_{r}$ on local ionic concentration, 
near surfaces, or when subjected to external fields \cite{hess2006modeling,bonthuis2012unraveling,schlaich2016water}. 
At sufficiently large $\varepsilon_{r}$ and sufficiently small bulk salt concentration $\rhos$, the essential physics 
is captured by a mean-field description in which correlations, image-charge interactions, and (in-plane) ordering are neglected \cite{torrie1979monte}.
The electrodes are assumed to extend to infinity to facilitate a description in which
physical quantities depend only on the coordinate $x$ perpendicular to the electrode surfaces.  
For simplicity, we consider the case without Stern layers (Appendix~\ref{appendix:stern} discusses their effect).

The initially homogenous electrolyte is exposed to a suddenly applied potential difference $2\Psi$ over the two electrodes, after which EDLs form near the electrode surfaces.
The local dimensionless electrostatic potential $\phi$,
related to the local electrostatic potential via multiplication with the thermal voltage $\kbt/e$ (with $k_{\rm B}$ Boltzmann's constant and $e$ the proton charge)
is governed by Poisson's equation (in SI units)
\begin{eqnarray}\label{eq:Poisson}
\partial_{x}^2 \phi=-4\pi \lb q,
\end{eqnarray}
with $\lb=e^{2}/(4\pi\varepsilon_{0}\varepsilon_{r} \kbt)$ being the Bjerrum length 
 and $\varepsilon_{0}$ being the vacuum permittivity, respectively. 
Moreover, $q$ is the reduced ionic charge density (unit m$^{-3}$), the difference between cationic and anionic number densities,
that is governed by a continuity equation,
\begin{eqnarray}\label{eq:continuity_ions}
\frac{\partial q}{\partial t}=-\partial_{x} I,
\end{eqnarray}
with $I$ being the reduced ionic current density (unit m$^{-2}$ s$^{-1}$), the difference between cationic and anionic current densities. 
From the reduced quantities $q$ and $I$ we find the ionic charge density and ionic current density as $qe$ and $Ie$, respectively.
For brevity, however, from hereon we omit the adjective ``reduced'' and speak simply of the ionic charge density $q$ and the ionic current density $I$.

Depending on the applied dimensionless electrode potential $\Phi\equiv e\Psi/\kbt$, different theories can be employed to obtain expressions 
for $I$.
For instance, 
the classical Nernst-Planck equations are applicable to dilute electrolytes up to roughly the thermal voltage 
$\Phi=1$.
Beyond this value, steric hinderance among ions, especially near electrode surfaces where ions can form layered packings \cite{hartel2017structure}, 
must be incorporated via, e.g., mean-field modifications \cite{kilic2007stericII} or dynamical density functional theory \cite{jiang2014time}.
The opposite limit of small applied potentials $\Phi\ll1$ gives rise to the Debye-Falkenhagen approximation in which the sum of 
locally-varying cationic and anionic densities is roughly $2\rhos$.
With this approximation, and assuming the same diffusion constant $D$ for both ion species,
one easily derives (see, e.g., Ref.~\cite{bazant2004diffuse}) the ionic current density,
\begin{eqnarray}\label{eq:ioniccurrent}
I=-D\left[\partial_{x}q+2\rhos\partial_{x}\phi\right],
\end{eqnarray}
from the Nernst-Planck equations for the individual ion species.
Henceforth we moreover assume $D$ to 
be independent of the local ionic concentrations. As we have assumed the temperature $T$ to be constant, \eqr{eq:ioniccurrent} does not contain a thermodiffusive term.

Combining Eqs.~\eqref{eq:Poisson}, \eqref{eq:continuity_ions}, and~\eqref{eq:ioniccurrent} gives rise to the Debye-Falkenhagen equation \cite{debye1928dispersion},
\begin{align}\label{eq:DebyeFalkenhagen}
\frac{\partial q}{\partial t}
=D\left[\partial_{x}^2 q-\kappa^{2} q\right],
\end{align}
with $\kappa=\ld^{-1}=\sqrt{8\pi \rhos \lb}$ being the inverse Debye length. 
The main task of this article is to determine the transient EDL formation arising from this equation. 
However, as \eqr{eq:DebyeFalkenhagen} solely captures ionic drift and diffusion, it cannot be expected to be reliable
on timescales where vibrations and rotations of individual molecules come into play.

To progress, we apply a Laplace transform on the time domain, which transforms a function $f(t)$ into $\hat{f}(s)\equiv\int_{0}^{\infty}f(t)\exp{[-ts]}dt$. We find
\begin{align}\label{eq:laplacechargedensity0}
\partial_{x}^2 \hat{q}&=k^2\hat{q}(x,s)-\frac{q(x,0)}{D},
\end{align}
with $k^2=\kappa^{2}+s/D$.  
For an initially homogenous electrolyte, $q(x,0)=0$, the antisymmetric solution [$\hat{q}(x)=-\hat{q}(-x)$] to 
\eqr{eq:laplacechargedensity0} reads
\begin{align}\label{eq:laplacechargedensity2}
\hat{q}(x,s)&=A_{1}\sinh(k x), 
\end{align}
with $A_{1}$ an integration constant to be determined. Inserting \eqr{eq:laplacechargedensity2} into \eqr{eq:Poisson} 
and integrating once yields
\begin{align}\label{eq:partialpsi}
-\partial_{x}  \hat{\phi}&=4\pi \lb\left[\frac{A_{1}}{k}\cosh(k x)+A_{2}\right].
\end{align}
With Eqs.~\eqref{eq:laplacechargedensity2} and \eqref{eq:partialpsi}, we then find the Laplace-transformed ionic current density,
\begin{align}\label{eq:currentA}
-\frac{\hat{I}}{D}=&\frac{A_{1}}{k}\left(k^2-\kappa^{2}\right) \cosh(k x)-A_{2}\kappa^2.
\end{align}
Imposing the ionic current density to vanish at the boundaries, $\hat{I}(\pm L,s)=0$, yields $A_{2}=A_{1}s\cosh(k L)/(k\kappa^2 D)$;  
hence, we find
\begin{align}\label{eq:currentA2}
\hat{I}=A_{1}\frac{s}{k}  \left[ \cosh(k L)-\cosh(k x)\right].
\end{align}
The electric field $\hat{E}=-\kbt\partial_{x}\phi/e$ now follows from \eqr{eq:partialpsi},
\begin{align}\label{eq:potentialgradient}
\frac{e \hat{E}}{\kbt}=&4\pi \lb \frac{A_{1}}{k}\left[\cosh(k x)+  \cosh(k L)\frac{s}{\kappa^{2}D}\right].
\end{align}
Integrating \eqr{eq:potentialgradient} gives the dimensionless potential,
\begin{align}\label{eq:fixingA}
-\hat{\phi}=&4\pi \lb \frac{A_{1}}{k^{2}}\left[\sinh(k x)+ \cosh(k L)\frac{s}{\kappa^{2}D}kx\right],
\end{align}
where the integration constant of this integration is zero due to antisymmetry of $\hat{\phi}$. 
$A_{1}$ is now fixed by the imposed time-varying surface potential 
$\phi(x=-L,t\ge0)=\Phi$. 
Its Laplace transform, $\hat{\phi}(x=-L,s)=\Phi/s$, is inserted in \eqr{eq:fixingA} to find 
\begin{align}\label{eq:Aconstant}
A_{1}&=\frac{\Phi}{4\pi \lb }\frac{k^2}{s}\frac{1}{\displaystyle{\sinh(k L)+  \cosh(k L)\frac{s}{\kappa^{2}D}kL}}.
\end{align}
This corresponds to Eq.~(26) of Ref.~\cite{bazant2004diffuse} (for a vanishing Stern layer width, $\ls=0$, and  
with a minus sign difference since Ref.~\cite{bazant2004diffuse} applies the opposite potentials at $x=\pm L$). 

\section{Pad\'{e} approximation before inverse Laplace transformation}\label{sec_pade}
In order to find the real-time response of an EDLC, at this point, 
previous authors \cite{bazant2004diffuse, stout2017diffuse, feicht2016discharging} 
choose to apply Pad\'{e} approximations to functions such as the ionic charge density.
The general spirit is to approximate a function $\hat{g}(s)$ around $s=\bar{s}$ by a rational function of the form 
\begin{equation}\label{eq:padeapproximation}
\hat{g}^{pq}(s)=\frac{\alpha_{0}+\alpha_{1}(s-\bar{s})+..+\alpha_{p}(s-\bar{s})^{p}}{\beta_{0}+\beta_{1}(s-\bar{s})
+..+\beta_{q}(s-\bar{s})^{q}}.
\end{equation}
The inverse Laplace transform to the approximated function $\hat{g}^{pq}(s)$ can then be readily performed to obtain $g^{pq}(t)$.

To get a feeling for the appropriateness of this method, 
consider the function  $g(t)=\textrm{erf}(\sqrt{t})$ for which $\hat{g}(s)=1/(s\sqrt{s+1})$. 
A low-order Pad\'{e} approximation for $\hat{g}(s)$ around $\bar{s}=0$ 
reads for instance $\hat{g}^{02}(s)=1/s-1/(2+s)$, which yields $g^{02}(t)=1-e^{-2t}$. 
Analogously one finds $g^{13}(t)=1-\frac{1}{2}\left(\exp{[-2(2+\sqrt{2})t]}+\exp{[-2(2-\sqrt{2})t]}\right)$. 
Reference~\cite{heavilin2012approximation} notes that both $g^{02}(t)$ and $g^{13}(t)$ approximate $g(t)$ fairly well. 
Caution should be taken however, if we are interested in the long-time relaxation of $g(t)$, $g(t\to\infty)\sim \exp[-t]/\sqrt{t}$. 
Clearly, $g^{02}(t)$ overestimates the relaxation by a factor 2. $g^{13}(t)$ does better with an overestimation by a factor 1.17.

Regarding our physical system of interest (an EDLC subject to a potential step), 
 such Pad\'{e} approximations give rise to questionable results.  
Consider, for example, the local charge density [\eqr{eq:laplacechargedensity2}]. 
This function has a pole at $s=0$, corresponding to the long-time limit of 
$q(x,t)$ [cf.~\eqr{eq:chargescontribution}], and an infinite amount of poles
on the negative real $s$ axis [cf.~Sec.~\ref{sec:poles}].
If we choose to apply a Pad\'{e} approximation on \eqr{eq:laplacechargedensity2} 
around $s/(D\kappa^2)=0$, then we find
\begin{subequations}
\begin{align}
\hat{q}^{02}(x,s)&=2\rhos \Phi \frac{\sinh(\kappa x)}{\sinh(\kappa L)}\frac{1}{s[1+\tau_{q}(x)s]}\label{eq:padechargedensity3},\\
\tau_{q}(x)&\equiv\frac{\ld L}{2D}  \left[\frac{3}{\tanh(\kappa L)}-\frac{x}{L\tanh(\kappa x)}- \frac{2\ld}{L}\right].\label{eq:padechargedensity4}
\end{align}
\end{subequations}
The inverse Laplace transform of $\hat{q}^{02}(x,s)$ then reads
\begin{align}\label{eq:padeionicchargedensity}
q^{02}(x,t)&=2\rhos \Phi \frac{\sinh(\kappa x)}{\sinh(\kappa L)}\left(1-\exp{\left[-\frac{t}{\tau_{q}(x)}\right]}\right).
\end{align} 
We note that, for $x=-L$, \eqr{eq:padechargedensity4} corresponds to Eq.~(30) of Ref.~\cite{bazant2004diffuse} 
for the case of vanishing Stern layer thickness. For future reference we report $\tau_{q}(-L)$ for 
limiting cases of $\kappa L$,
\begin{align}\label{eq:limitingtimescalesq}
 \tau_{q}(-L)&=
 \begin{cases}
       \displaystyle{\frac{\ld L}{D}  \left[1-\frac{1}{\kappa L}+\mathcal{O}\left(\exp[-2\kappa L]\right)\right]},& \hspace{0.1cm}\kappa L\gg1, \\ 
     \displaystyle{\frac{L^2}{D}  \left[\frac{1}{3}+\frac{1}{45}(\kappa L)^2+\mathcal{O}\left((\kappa L)^4\right)\right]},& \hspace{0.1cm} \kappa L\ll1.
    \end{cases}
\end{align}
As the approximated function $\hat{q}^{02}(x,s)$ is most accurate around $s/(D\kappa^2)=0$ (the point around which we expanded), 
we find that, at long times, \eqr{eq:padeionicchargedensity} correctly relaxes to the Debye-H\"{u}ckel ionic charge density. 
The first pole $s_{1}$ of $\hat{q}(x,s)$ that one encounters departing from $s/(D\kappa^2)=0$, 
i.e., the pole with the largest (least negative) real part, determines the long-time relaxation of $q(x,t)$.
Clearly, the accuracy of the Pad\'{e} approximation $\hat{q}^{02}(x,s\approx s_{1})$ around that pole depends on its distance from $s/(D\kappa^2)=0$.

Remarkably, while the pole structure of $\hat{q}(x,s)$ does not depend on $x$, 
the pole structure of its Pad\'{e} approximation $\hat{q}^{02}(x,s)$ does, 
leading to an $x$-dependent decay time $\tau_{q}$. 
Ultimately, this $x$ dependence arises because, in determining the coefficients $\alpha_{0},..,\alpha_{p},\beta_{0},..,\beta_{q}$ 
of the Pad\'{e} approximation, a linear system of $p+q+2$ equations has to be solved,
which acquire $x$ dependence from the numerator in \eqr{eq:laplacechargedensity2}.
Consequently, analogous approximations to the the ionic current density and electric field exhibit decay times
$\tau_{I}$ and $\tau_{E}$ with $\tau_{q}\neq\tau_{I}\neq\tau_{E}$ \cite{Stoutfootnote}.
However, it follows from Eqs.~\eqref{eq:Poisson} and \eqref{eq:continuity_ions} that all these timescales should be equal. 
Physically speaking, 
compared to the ionic dynamics, the electromagnetic field readjusts itself instantaneously to always follow suit [\eqr{eq:Poisson}], 
and changes in ionic charge density cannot be faster or slower 
than the concomitant ionic current densities [\eqr{eq:continuity_ions}].
In order to avoid such unphysical features as $x$-dependent relaxation times or different 
timescales for related quantities, which will persist regardless of the chosen orders $p$ and $q$, we will not rely on Pad\'{e} approximation schemes in the following.

\section{Exact inverse Laplace transformation}\label{sec:inverse} 
We report expressions for the inverse Laplace transforms of the ionic charge density [\eqr{eq:laplacechargedensity2}], the ionic 
current density [\eqr{eq:currentA2}], 
and the electric field [\eqr{eq:potentialgradient}]. 
To tidy up our notation, we introduce $m\equiv kL$ and $n\equiv \kappa L$, allowing us to rewrite  
$s=(m^{2}-n^{2})D/L^2$ and 
\begin{align}\label{eq:A1}
A_{1}&= \frac{\bar{q} m}{s}\hat{f}(m,n), 
\end{align}
with $\bar{q}\equiv\Phi/(4\pi \lb L^2)$ and 
\begin{align}\label{eq:denominator}
\hat{f}(m,n)\equiv\left[\frac{\sinh m}{m}+ \left[\frac{m^2}{n^2}-1\right] \cosh m\right]^{-1}.
\end{align}

\subsection{Ionic charge density}\label{subsec:ionic}
In terms of these new variables, the ionic charge density [\eqr{eq:laplacechargedensity2}] reads
\begin{align}\label{eq:localchargedensity}
\hat{q}(x,s)&=\frac{\bar{q}m}{s}\hat{f}(m,n)\sinh\frac{m\kappa x}{n}.
\end{align}
Obtaining $q(x,t)$ requires evaluating a Bromwich integral 
\begin{align}\label{eq:laplacebacktransformdefinition}
q(x,t)=&\frac{1}{2\pi i}\oint_{\gamma}\exp{[st]}\hat{q}(x,s)ds\nn
=&\sum_{\ell}\text{Res}\left(\exp{[st]}\hat{q}(x,s), s_{\ell}\right),
\end{align}
with $s, s_{\ell}\in\mathbb{C}$ and $\ell$ enumerating the poles $s_{\ell}$ of $\hat{q}(s)$. Moreover, 
 $\gamma$ is a path that consists of the line from $c-i\infty$ to $c+i\infty$, with $c\in\mathbb{R}$ such that
 $c>\text{Re}(s_{\ell})$ for all $\ell$, together with a semi-circle that encloses all poles $s_{\ell}$.

Besides the pole $s_{0}\equiv0$, the poles of $\hat{q}$ coincide with the poles $s_{j}$ of the term $\hat{f}(m,n)$; 
hence, $s_{\ell}=\{s_{0},s_{j}\}$.
The pole $s_{0}$ gives rise to the contribution 
\begin{align}\label{eq:chargescontribution}
\text{Res}\left(\exp{[st]}\hat{q}(s), s_{0}\right)&=\lim_{s\to 0}\left[\bar{q}m\sinh \left(\frac{m\kappa x}{n}\right)\hat{f}(m,n)\right]\nn
&=2\rhos\Phi\frac{\sinh (\kappa x)}{\sinh n}
\end{align}
to \eqr{eq:laplacebacktransformdefinition},
where we used that $s=0 \Leftrightarrow m=n$ and $\hat{f}(n,n)=n/\sinh n$. 
To determine the contributions of the poles $s_{j}$ to \eqr{eq:laplacebacktransformdefinition}, 
we need to determine the locations of these poles.

\subsubsection{Poles of $\hat{f}(n,m)$}\label{sec:poles}
Finding the pole locations $s_{j}$ boils down to determining the solutions to the transcendental equation
\begin{align}\label{eq:mpole}
\tanh m= m \left(1-\frac{m^2}{n^2}\right), \hspace{1cm} m\in\mathbb{C}.
\end{align}
By means of a systematic numerical investigation, we expect there to be no solutions to \eqr{eq:mpole}
other than those that lie on the real or imaginary $m$ axes.
In what follows we thus 
consider either $m=\tilde{m}\in \mathbb{R}$, for which we need to solve 
\begin{align}\label{eq:realmpole}
\tanh \tilde{m}= \tilde{m} \left(1-\frac{\tilde{m}^2}{n^2}\right), \hspace{1cm} \tilde{m}\in\mathbb{R},
\end{align}
or $m=iM$, $M\in \mathbb{R}$, for which we need to solve 
\begin{align}\label{eq:imaginarympole}
\tan M&=M\left(1+\frac{M^2}{n^2}\right), \hspace{1cm} M\in\mathbb{R}.
\end{align}

\begin{figure}[h!]
\centering
\includegraphics[width=0.48\textwidth]{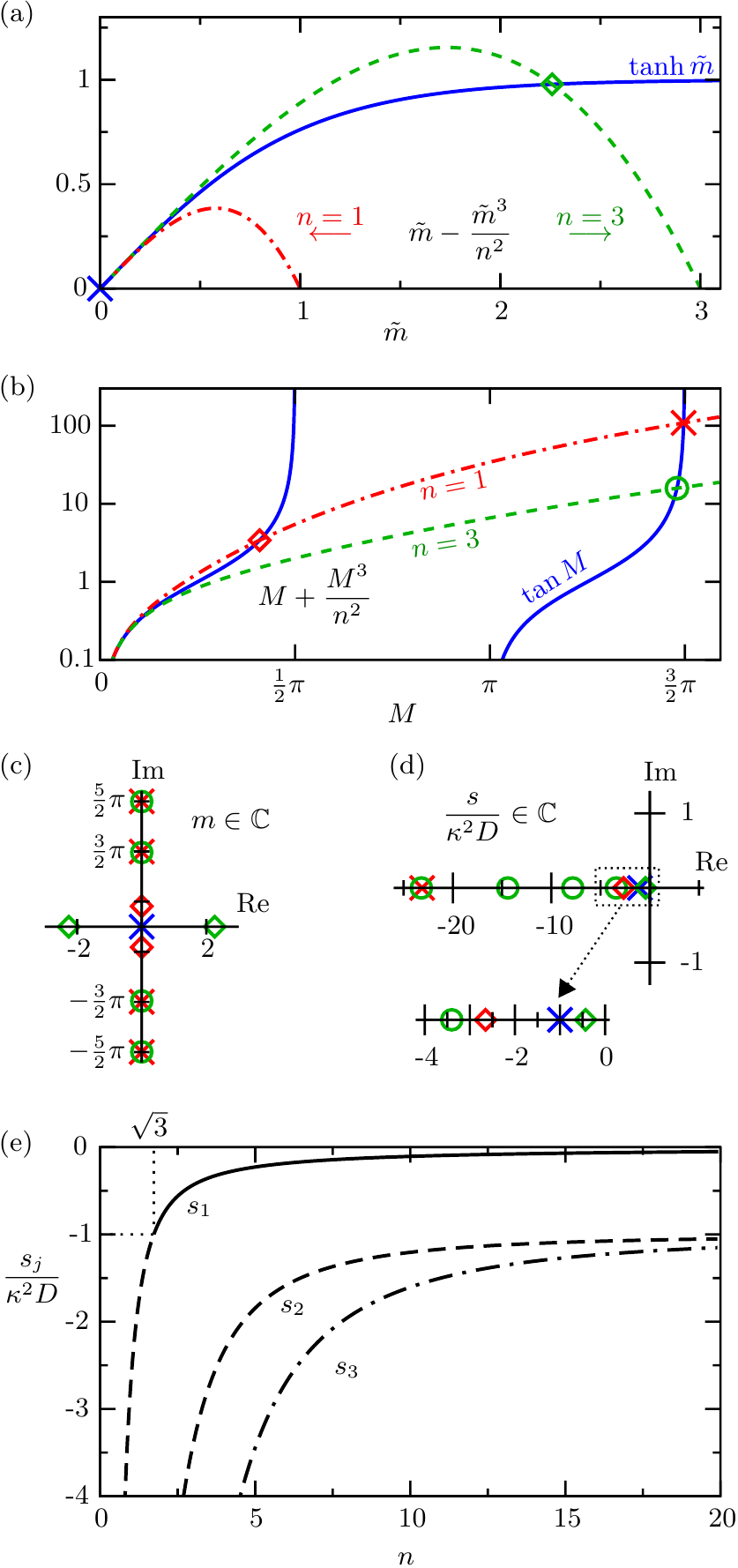}
  \caption{Solutions to \eqr{eq:mpole} are found on the real (a) 
	  and imaginary (b) $m$ axes as the intersections in these graphs. The solid blue lines indicate 
	  $\tanh m$ (a) and $\tan m$ (b), respectively. The other lines indicate the polynomials 
	  $m\pm m^3/n^2$ at $n=1$ (red dash-dotted) and $n=3$ (green dashed). 
	  The solutions found in (a) and (b) are portrayed in the complex $m\in \mathbb{C}$ (c)
	   and $s/(\kappa^2 D)\in \mathbb{C}$ (d) planes.
	  (e) The $n$ dependence of $s_{j}$ for $j=\{1,2,3\}$ (dashed/solid, dashed, dash-dotted).}
    \label{fig:poles_of_f}
\end{figure}

In Fig.~\ref{fig:poles_of_f} we show the left-hand side (solid blue line) and the right-hand side (dashed green line and dash-dotted red line) 
of \eqr{eq:realmpole}~[Fig.~\ref{fig:poles_of_f}(a)] 
and \eqr{eq:imaginarympole}~[Fig.~\ref{fig:poles_of_f}(b)], respectively. The intersections of these lines indicate solutions to the equations. 
First, due to the periodic nature of $\tan M$, we find an infinite amount of solutions to \eqr{eq:imaginarympole}, 
which we denote $\pm M_{j}$ where $j\in \mathbb{N}$ labels the pole that lies in the interval $(j-1)\pi<M_{j}<(j-1/2)\pi$.
While the poles at $M_{j\ge2}$ are present regardless of the value of $n>0$,
there exists a nontrivial solution $0<M_{1}<\pi/2$ to \eqr{eq:imaginarympole} but no solution $\tilde{m}_{1}$ to \eqr{eq:realmpole}
in the case $n<\sqrt{3}$, whereas the opposite situation occurs in the case $n>\sqrt{3}$.
This behavior is summarized in Fig.~\ref{fig:poles_of_f}(c). There, also the trivial solution ($m_{0}\equiv0$) to \eqr{eq:mpole} is shown.

For future convenience, we introduce the symbol $\mathcal{M}_{j}$, 
\begin{align}\label{eq:mathcalM}
\mathcal{M}_{1}=&
\begin{cases}
      M_{1}& ,\hspace{0.5cm}n<\sqrt{3}, \\
      i\tilde{m}_{1}& ,\hspace{0.5cm} n>\sqrt{3},\nn
\end{cases}\\
\mathcal{M}_{j\ge2}=& M_{j},
\end{align}
with $j\in\mathbb{N}$, and
with $\tilde{m}_{1}$ and $M_{j}$ the solutions to \eqr{eq:realmpole} and \eqr{eq:imaginarympole}, respectively.
The following table summarizes $\mathcal{M}_{j}$ for various values of $n$: 
\begin{center}
    \begin{tabular}{ | l | l | l | l | l | l |}
    \hline
      & $\mathcal{M}_{1}$ & $\mathcal{M}_{2}$ & $\mathcal{M}_{3}$ &$\mathcal{M}_{4}$ &$\mathcal{M}_{5}$\\
    \hline
    $n=10$          & $i9.456$& 4.531 & 7.774 & 10.954 & 14.114\\
    $n=3$           & $i2.259$& 4.649 & 7.838 & 10.989 & 14.134\\
    $n=\sqrt{3}$ 	  &0       & 4.687 & 7.848 & 10.993 & 14.136\\ 
    $n=1$           & 1.286   & 4.703 & 7.852 & 10.995 & 14.137\\ 
    $n=0.1$         & 1.568   & 4.712 & 7.854 & 10.996 & 14.137\\\hline \hline
    $(2j-1)\pi/2$ & 1.571   & 4.712 & 7.854 & 10.996 & 14.137\\
    \hline
    \end{tabular}
\end{center}
For small $n$, the deviation $\epsilon$ of $M_{1}$ from $\pi/2$ is found by inserting $M_{1}=\pi/2-\epsilon$ 
into \eqr{eq:imaginarympole}, which gives $\epsilon=8n^2/\pi^3$.
We find $M_{1}=1.568$ for $n=0.1$, in accordance with the numeric solution.
For small $n$, the same arguments lead to the same corrections to $\mathcal{M}_{j\ge2}=(2j-1)\pi/2+\mathcal{O}\left(n^{2}\right)$.

Regarding the solution $\tilde{m}_{1}$, it is clear from inspection of Fig.~\ref{fig:poles_of_f}(a) that $\tilde{m}_{1}$ 
increases with $n$. For large $n$, $\tanh\tilde{m}_{1}\approx1$ hence 
$\tilde{m}_{1}$ is the solution to $1=\tilde{m}_{1}(1-\tilde{m}_{1}^2/n^2)$. 
From this we infer that at large $n$ the fraction $\tilde{m}_{1}/n\to1$. 
Setting $z\equiv\tilde{m}_{1}/n-1$ which is a solution to $1/n=(z+1)(z+2)(-z)=-2z-3z^2-z^3$, one obtains
$z=-\left(1/n+3z^2+z^3\right)/2=-1/(2n)+\mathcal{O}\left(n^{-2}\right)$
and hence, the solution $\tilde{m}_{1}$, present if $n>\sqrt{3}$, is approximated by 
\begin{align}\label{eq:mtilde}
\tilde{m}_{1}&=n+nz=n-\frac{1}{2}+\mathcal{O}\left(n^{-1}\right).
\end{align}
Indeed, the table above shows that $\tilde{m}_{1}\approx 9.5$ at $n=10$.

Given the definition $m=L\sqrt{\kappa^2+s/D}$, two poles in $m\in \mathbb{C}$ correspond to one pole in $s\in \mathbb{C}$.
In particular, the sets of poles $\pm \tilde{m}_{1}$ and $\pm i M_{j}$ correspond to poles at
$s=-D(\kappa^{2}-\tilde{m}_{1}^{2}/L^{2})$ and $s=-D(\kappa^{2}+M_{j}^{2}/L^{2})$, respectively [see Fig.~\ref{fig:poles_of_f}(d)]. 
Using the symbol $\mathcal{M}_{j}$ as defined in \eqr{eq:mathcalM}, the locations of the poles in $s\in \mathbb{C}$ are given by
\begin{equation}\label{eq:sj}
s_{j}=\displaystyle{-D\kappa^{2}\left(1+\frac{\mathcal{M}_{j}^{2}}{n^{2}}\right)}.
\end{equation}
While we found a transition at $n=\sqrt{3}$ from an $M_{1}$ to an $\tilde{m}_{1}$ solution in $m\in\mathbb{C}$,
we find no special behavior in the pole structure of $s\in\mathbb{C}$ at that point. 
As is clear from \eqr{eq:laplacebacktransformdefinition}, the locations of the poles $s_{j}$ 
determine the temporal behavior of $q(x,t)$. In particular, 
these poles satisfy $\text{Im}(s_{j})=0$ and $\text{Re}(s_{j})<0$. The latter property,
which ensures that $q(x,t)$
decays monotonically over time, is obvious for $j\ge2$, and for $j=1$ in the case $n<\sqrt{3}$. 
At any finite $n>\sqrt{3}$, the property $s_{1}<0$ follows from \eqr{eq:realmpole}:
\begin{align}
&\frac{\tilde{m}_{1}^2}{n^2}=1-\frac{\tanh \tilde{m}_{1}}{\tilde{m}_{1}} \in [0,1) \nn
&\Rightarrow \kappa^2+\frac{s_{1}}{D}<\kappa^{2}\Rightarrow s_{1}<0.
\end{align}
The poles $s_{j}$ have the dimension of inverse time; hence, give rise to timescales $\tau_{j}\equiv-1/s_{j}$:
\begin{equation}\label{eq:tauj}
\tau_{j}=\frac{L^{2}}{\displaystyle{D\left(n^{2}+\mathcal{M}_{j}^{2}\right)}},
\end{equation}
which are not only the characteristic relaxation timescales of the ionic charge density [cf.~\eqr{eq:exactq}], 
but also of the ionic current [cf.~\eqr{eq:exactcurrent}] and the electric field [cf. \eqr{eq:exactEfield}].
The pole $s_{1}$ with the largest (i.e., least negative) real part, which determines the slowest decay mode (largest $\tau_{j}$), 
is displayed in Fig.~\ref{fig:poles_of_f}(e) as a function of $n=\kappa L$.
For $n\gg1$, $|s_{1}|$ becomes small, which a posteriori justifies the 
expansion around $s/(\kappa^2 D)=0$ underlying the Pad\'{e} approximation schemes cited in Sec.~\ref{sec_pade}.
However, $s_{1}\to-\infty$ for strongly overlapping double layers ($n\ll1$); hence, a Pad\'{e} approximation  
 around $s/(\kappa^2 D)=0$ of $\hat{q}(x,s)$ might not approximate $\hat{q}(x,s)$ around $s_{1}$ equally accurately.

\subsubsection{Residues of $\hat{q}(x,s)$ at $s_{\ell}$}
In the vicinity of $\pm\tilde{m}_{1}$ we find 
\begin{align}\label{eq:fnearm}
\frac{1}{\hat{f}(m,n)}&\displaystyle{\overset{m\to \pm \tilde{m}_{1}}=}\pm\frac{1}{A_{\tilde{m}_{1}}}(m\mp \tilde{m}_{1})+\mathcal{O}\left((m\mp \tilde{m}_{1})^{2}\right),\nn
\hat{f}(m,n)&\overset{m\to \pm \tilde{m}_{1}}{=} 
\pm\frac{ A_{\tilde{m}_{1}}}{m\mp \tilde{m}_{1}}+\mathcal{O}\left((m\mp \tilde{m}_{1})^{0}\right),
\end{align}
with
\begin{align}\label{eq:A}
 A_{\tilde{m}_{1}}\equiv\frac{\tilde{m}_{1}^{2}}{\displaystyle{\left[\tilde{m}_{1}+\frac{2\tilde{m}_{1}^3}{n^2}\right]\cosh \tilde{m}_{1}-\left[1+\tilde{m}_{1}^2-\frac{\tilde{m}_{1}^4}{n^2}\right]\sinh \tilde{m}_{1}}}.
\end{align}
Similarly, in the vicinity of $\pm i M_{j}$ we find
\begin{align}\label{eq:fnearM}
\hat{f}(m,n)&\overset{m\to \pm iM_{j}}{=} 
\pm \frac{iA_{M_{j}}}{(m\mp iM_{j})}+\mathcal{O}\left((m\mp iM_{j})^{0}\right),
\end{align}
with 
\begin{align}
A_{M_{j}}\equiv\frac{M_{j}^2}{\displaystyle{\left[M_{j}-\frac{2M_{j}^3}{n^2}\right]\cos M_{j} -\left[1-M_{j}^2-\frac{M_{j}^4}{n^2}\right]\sin M_{j}}}. 
\end{align}
The second terms on the right-hand sides of Eqs.~\eqref{eq:fnearm} and \eqref{eq:fnearM} 
contain no poles and hence do not contribute to \eqr{eq:laplacebacktransformdefinition}.

Noting that the poles of $\hat{f}(m,n)$ occur in pairs, we can 
consider the sum of the poles at $\pm\tilde{m}_{1}$ of
$\hat{f}(m,n)$, 
\begin{align}
\frac{A_{\tilde{m}_{1}}}{m- \tilde{m}_{1}}-\frac{A_{\tilde{m}_{1}}}{m+\tilde{m}_{1}}
&=\frac{2A_{\tilde{m}_{1}}\tilde{m}_{1}}{\displaystyle{n^2+\frac{s L^2}{D}- \tilde{m}_{1}^{2}}},
\end{align}
and the sum of poles at $iM_{j}$ and $-iM_{j}$, 
\begin{align}
\frac{iA_{M_{j}}}{m- iM_{j}}-\frac{iA_{M_{j}}}{m+iM_{j}}
&=-\frac{2A_{M_{j}}M_{j}}{\displaystyle{n^2+\frac{s L^2}{D}+M_{j}^2}}.
\end{align}
Hence, two poles at $\pm \tilde{m}_{1}$ contribute a single pole, 
\begin{align}\label{eq:fnearm1}
\hat{f}^{\tilde{m}_{1}}(s)&=\frac{2DA_{\tilde{m}_{1}}\tilde{m}_{1}}{L^2}\frac{1}{s-s_{1}},
\end{align}
located in $s\in\mathbb{C}$ at $s_{1}=-\left[D\left(\kappa^{2}-\tilde{m}_{1}^2/L^2\right)\right]$, 
and two poles located at $\pm iM_{j}$ contribute a single pole,
\begin{align}
\hat{f}^{M_{j}}(s)&=-\frac{2DA_{M_{j}}M_{j}}{L^2}\frac{1}{s-s_{j}},
\end{align}
at $s_{j}=-D\left[\kappa+M_{j}^2/L^2\right]$ to the sum in \eqr{eq:laplacebacktransformdefinition}. The solution $m_{0}=0$ to \eqr{eq:mpole} does not contribute to this sum as its residue is zero.

For $n>\sqrt{3}$, the pole $s_{1}$ gives
\begin{widetext}
\begin{align}\label{eq:chargemcontribution}
\text{Res}\left(\hat{q}(s)\exp{[st]}, s_{1}\right)
&\overset{n>\sqrt{3}}=\frac{\bar{q}L^2}{D}\lim_{s\to s_{1}}\left[(s-s_{1})\frac{m}{m^{2}-n^{2}}\sinh\left(\frac{m\kappa x}{n}\right)\hat{f}^{\tilde{m}_{1}}(s)\exp{[st]}\right]\\
&\overset{\phantom{n>\sqrt{3}}}=\frac{2\bar{q}}{\tilde{m}_{1}^2-n^2} 
\frac{\displaystyle \tilde{m}_{1}^4 \sinh \frac{\tilde{m}_{1}\kappa x}{n}}{\displaystyle\left[\tilde{m}_{1}+
\frac{2\tilde{m}_{1}^3}{n^2}\right]\cosh \tilde{m}_{1} - \left[1+\tilde{m}_{1}^2
-\frac{\tilde{m}_{1}^4}{n^2}\right]\sinh \tilde{m}_{1} }\exp{\left[-\frac{D\left(n^{2}-\tilde{m}_{1}^{2}\right)t}{L^2}\right]}, \nonumber
\end{align}
where we used $s=(m^2-n^2)D/L^2$ to obtain the first line from \eqr{eq:localchargedensity} and, going to the second line, we used Eqs.~\eqref{eq:fnearm1} and \eqref{eq:A}.
For $n<\sqrt{3}$, the poles $s_{j}$ give
\begin{align}\label{eq:chargeMcontribution}
\sum_{j\ge1}\text{Res}\left(\hat{q}(s)\exp{[st]}, s_{j}\right)
&\overset{n<\sqrt{3}}=\frac{\bar{q}L^2}{D}\sum_{j\ge1}
\lim_{ s\to s_{j}}\left[(s-s_{j})\frac{m}{m^{2}-n^{2}}\sinh \left(\frac{m\kappa x}{n}\right)\hat{f}^{M_{j}}(s)\exp{[st]}\right]\\
&\overset{\phantom{n<\sqrt{3}}}=-
\sum_{j\ge1}
  \frac{2\bar{q}}{M_{j}^2+n^2}
  \frac{\displaystyle M_{j}^{4}\sin \frac{M_{j}\kappa x}{n}}
  {\displaystyle \left[M_{j}-\frac{2M_{j}^3}{n^2}\right]\cos M_{j}- \left[1-M_{j}^2-\frac{M_{j}^4}{n^2}\right]\sin M_{j} }
\exp{\left[-\frac{D\left(n^2+M_{j}^2\right)t}{L^2}\right]},\nonumber
\end{align}
while for $n>\sqrt{3}$, the term $\hat{f}^{M_{1}}$ is absent and the above sums start at $j=2$. 
We can now conveniently write \eqr{eq:chargemcontribution} and
\eqr{eq:chargeMcontribution} as a single equation by replacing $M_{j}$ by $\mathcal{M}_{j}$ [see \eqr{eq:mathcalM}] in \eqr{eq:chargeMcontribution}.
This replacement accounts for all poles $s_{j}$ regardless of the value of $n$.
Using \eqr{eq:laplacebacktransformdefinition} and \eqr{eq:tauj}, we find
\begin{align}\label{eq:exactq}
\frac{q(x,t)}{\bar{q}}=&n^2 \frac{\sinh (\kappa x)}{\sinh n}
-\sum_{j\ge1}\frac{1}{\mathcal{M}_{j}^2+n^2}
  \frac{\displaystyle{2 \mathcal{M}_{j}^{4} \sin \frac{\mathcal{M}_{j}\kappa x}{n}}}
  {\displaystyle{\left[\mathcal{M}_{j}-\frac{2\mathcal{M}_{j}^3}{n^2}\right]\cos \mathcal{M}_{j} -  \left[1-\mathcal{M}_{j}^2-\frac{\mathcal{M}_{j}^4}{n^2}\right]\sin \mathcal{M}_{j}}}
\exp{\left[-\frac{t}{\tau_{j}}\right]}.
\end{align}
%\end{widetext}

\subsection{Ionic current density}
Inserting $A_{1}$, the ionic current density $\hat{I}$ [\eqr{eq:currentA2}] reads
\begin{align}\label{eq:currentxs}
\hat{I}(x,s)=&\bar{q}L\left[\cosh m-\cosh \frac{m\kappa x}{n}\right]\hat{f}(m,n).
\end{align}
Hence, the ionic current density $\hat{I}$ has the same poles $s_{j}$ as the ionic charge density $\hat{q}$, but lacks the pole $s_{0}$. 
This means that the ionic current density decays to zero at long times 
 with the same timescales $\tau_{j}$ as the ionic charge density.
The current could again be computed with the residue theorem, but the same result (as we have checked) can be obtained via a short-cut
that uses \eqr{eq:continuity_ions} to write
$I(x,t)=\cancel{I(x=-L,t)}-\int_{-L}^{x}dx \partial_{t}q$. Inserting \eqr{eq:exactq}, we find
%\begin{widetext}
\begin{align}\label{eq:exactcurrent}
I(x,t)=&-\frac{2\bar{q}D}{L}
\sum_{j\ge1}\frac{\displaystyle \mathcal{M}_{j}^{3}\left[\cos \mathcal{M}_{j}-\cos \frac{\mathcal{M}_{j}
\kappa x}{n}\right]}{\displaystyle \left[\mathcal{M}_{j}-\frac{2\mathcal{M}_{j}^3}{n^2}\right]\cos \mathcal{M}_{j}- 
\left[1-\mathcal{M}_{j}^2-\frac{\mathcal{M}_{j}^4}{n^2}\right]\sin \mathcal{M}_{j} }\exp
\left[-\frac{t}{\tau_{j}}\right].
\end{align}
\end{widetext}

\subsection{Electric field}
Using \eqr{eq:A1}, we rewrite electric field [\eqr{eq:potentialgradient}] to 
\begin{align}\label{eq:Exs}
\hat{E}(x,s)=&\frac{\Psi}{L}\frac{\ld^2}{D}\left[\cosh m +\frac{\kappa^2 D}{s}\cosh \frac{m \kappa x}{n}\right]\hat{f}(m,n)\nn
\equiv&\hat{E}_{1}+\hat{E}_{2},
\end{align}
with $\Psi=\Phi \kbt/e$ being the surface potential (unit volt). 

The inverse Laplace transform $E_{1}$ of the first term $\hat{E}_{1}\sim \cosh m$ in \eqr{eq:Exs} is easily found: We can  
generalize our results for the ionic current density \eqr{eq:currentxs} where the same term appears with a different prefactor. 
We see that the prefactors of \eqr{eq:currentxs}
and \eqr{eq:exactcurrent} differ by $-2D/L^2$. Hence, we can find $E_{1}$ by selecting the  $\sim \cos \mathcal{M}_{j}$ 
term of the numerator of \eqr{eq:exactcurrent} and find the prefactor of $E_{1}$ 
by multiplying the prefactor of $\hat{E_{1}}$ [\eqr{eq:Exs}] with $-2D/L^2$.

The term $\hat{E}_{2}$ has the same poles $s_{\ell}=\{s_{0},s_{j}\}$ as
the ionic charge density [\eqr{eq:localchargedensity}]; hence,
\begin{align}\label{eq:Eresidues}
E_{2}&=\sum_{\ell}\text{Res}\left(\exp{[st]}\hat{E}_{2}(s), s_{\ell}\right).
\end{align}
The pole $s_{0}$ gives rise to  the Debye-H\"{u}ckel electric field, 
\begin{align}
\text{Res}\left(\exp{[st]}\hat{E}_{2}(s), s_{0}\right)
&=\frac{\Psi}{\ld}\frac{\cosh (\kappa x)}{\sinh n},
\end{align}
to which the electric field $E(t)$ relaxes at long times.

For $n<\sqrt{3}$, the poles $s_{j}$ give
\begin{widetext}
\begin{align}
\sum_{j\ge1}\text{Res}\left(\hat{E}_{2}(s)\exp{[st]}, s_{j}\right)
&\overset{n<\sqrt{3}}
=\frac{\Psi}{L}\sum_{j\ge1}
  \frac{2}{M_{j}^2+n^2}
  \frac{\displaystyle M_{j}^{3}\cos \frac{M_{j} \kappa x}{n}}
  {\displaystyle \left[M_{j}-\frac{2M_{j}^3}{n^2}\right]\cos M_{j}- \left[1-M_{j}^2-\frac{M_{j}^4}{n^2}\right]\sin M_{j} }
\exp{\left[-\frac{D\left(n^2+M_{j}^2\right)t}{L^2}\right]}.
\end{align}
Similarly to what was found for the ionic charge density [\eqr{eq:chargeMcontribution}], for the case $n>\sqrt{3}$, the term $j=1$ is absent and the above sum starts at $j=2$. 
A straightforward calculation now shows that we can again replace $M_{j}$ by $\mathcal{M}_{j}$ in the above equation to correctly 
capture the pole $s_{1}$ also for $n>\sqrt{3}$.
Putting everything together we find 
\begin{align}\label{eq:exactEfield}
E(x,t)=&\frac{\Psi}{\ld}\frac{\cosh (\kappa x)}{\sinh n}+\frac{\Psi}{L}\displaystyle{\sum_{j\ge1}\left[\frac{\displaystyle \cos \frac{\mathcal{M}_{j}\kappa x}{n}}
{\displaystyle\mathcal{M}_{j}^2+n^2}-\frac{\cos \mathcal{M}_{j}}{n^2}\right]
  \frac{2 \mathcal{M}_{j}^{3} }
  {\displaystyle\left[\mathcal{M}_{j}-\frac{2\mathcal{M}_{j}^3}{n^2}\right]\cos \mathcal{M}_{j} - \left[1-\mathcal{M}_{j}^2-\frac{\mathcal{M}_{j}^4}{n^2}\right]\sin \mathcal{M}_{j}}
\exp{\left[-\frac{t}{\tau_{j}}\right]}}.
\end{align}
\end{widetext}

\section{Discussion}\label{sec:discussion} 
\subsection{The timescales $\tau_{j}$}
\begin{figure}[h!]
\centering
\includegraphics[width=0.48\textwidth]{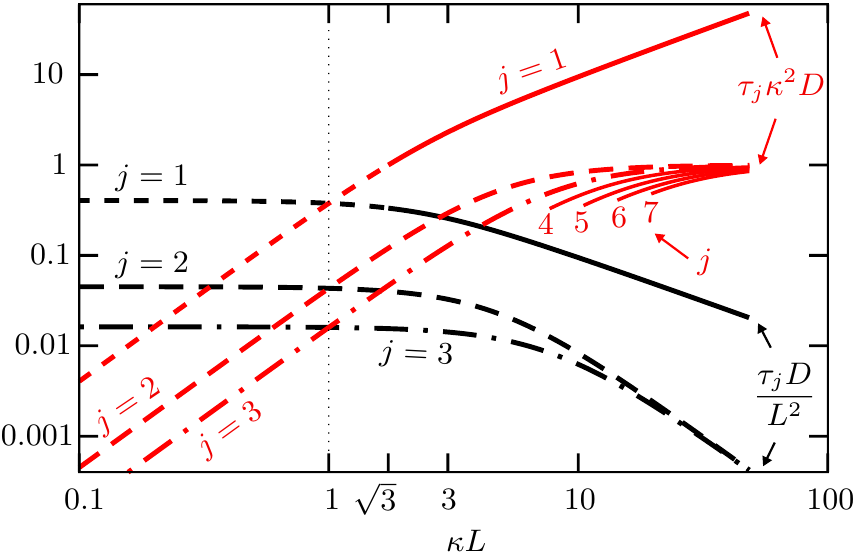}
 \caption{The decay time $\tau_{j}$ for several $j\le7$, nondimensionalized with $L^2/D$ and $1/(\kappa^2 D)$.}  
 \label{fig:tau1}
\end{figure}

The expressions for the ionic charge density [\eqr{eq:exactq}], ionic current density [\eqr{eq:exactcurrent}], 
and electric field [\eqr{eq:exactEfield}] all decay with the same timescales $\tau_{j}$ [\eqr{eq:tauj}].
Restoring conventional notation ($n\equiv \kappa L$), in Fig.~\ref{fig:tau1} 
we plot the $\kappa L$ dependence of the three largest timescales $\tau_{1}, \tau_{2}$, and $\tau_{3}$, 
where we use $L^2/D$ and $1/(\kappa^2 D)$, 
respectively, to nondimensionalize these timescales. Hence, at $\kappa L=1$ (dotted line), both ways of nondimensionalizing $\tau_{j}$ coincide. 
The behavior observed in Fig.~\ref{fig:tau1} is understood as follows. First, for $\kappa L\gg\sqrt{3}$,
\eqr{eq:mtilde} implies 
$\tilde{m}_{1}^{2}=(\kappa L)^{2}-\kappa L+\mathcal{O}\left((\kappa L)^{0}\right)$, which, filled in into \eqr{eq:tauj}, leads to 
\begin{align}\label{eq:tau1largen}
\tau_{1}&=\frac{L}{\kappa D}\left[1+\mathcal{O}\left(\frac{1}{\kappa L}\right)\right]\hspace{1.5cm}(\kappa L\gg\sqrt{3}),
\end{align}
confirming \eqr{eq:limitingtimescalesq} found via Pad\'{e} approximation. The high quality of this approximation
is understood with Fig.~\ref{fig:poles_of_f}(e)
which shows, in the limit $\kappa L\gg1$, that the pole $s_{1}$ approaches $s/(D\kappa^2)=0$, the point around which the Pad\'{e} approximation $\hat{q}^{02}(x,s)$
[\eqr{eq:padechargedensity3}] was performed. In that case, $\hat{q}^{02}(x,s)$ must also be accurate around $s_{1}$.

Notably, from the definition of the timescale $\tau_{j}$ [\eqr{eq:tauj}], 
at large $\kappa L$, many modes $j\ge2$ approach the Debye time [see Fig.~\ref{fig:tau1}]. Hence,
with increasing $\kappa L$, one needs an increasing amount of modes to accurately describe 
EDLC quantities around the Debye time. 

This collapse of timescales $\tau_{j\ge2}$ is not observed in the opposite limit of strongly overlapping double layers
($\kappa L\ll1$). 
Instead, in this limit, $\mathcal{M}_{j}=(2j-1)\pi/2+\mathcal{O}\left((\kappa L)^{2}\right)$, 
hence $\tau_{j}=4L^2/\left[D\left((2j-1)^2\pi^2\right)\right]+\mathcal{O}\left((\kappa L)^{2}\right)$,
which sets the heights of the plateaus observed in Fig.~\ref{fig:tau1}.
In particular, we find the long-time decay
\begin{align}\label{eq:tau1smalln}
\tau_{1}&=\frac{4L^2}{\pi^2 D}\left[1+\mathcal{O}\left((\kappa L)^{2}\right)\right]\hspace{1.2cm}(\kappa L\ll\sqrt{3}),
\end{align}
in agreement with the scaling found in \eqr{eq:limitingtimescalesq}. 
Importantly, as this article treats ionic dynamics via the mean-field Debye-Falkenhagen equation, 
small values $\kappa L\ll1$ cannot be reached by decreasing the electrode separation $2L$ down to the molecular size of the ions and solvent particles, 
but rather by electrolytes with low salt concentration, hence small inverse Debye lengths.

The factor $4/\pi^2=0.405$ in \eqr{eq:tau1smalln} constitutes a correction of 22\% over the factor $1/3$ in \eqr{eq:limitingtimescalesq}.
Our expression for the ionic charge density [\eqr{eq:exactq}], whose decay time is position independent, gives rise to a 
total ionic charge near one electrode [$Q(t)\equiv\int_{-L}^{0}q(x,t)dx$] that necessarily decays with the same relaxation time. 
Conversely, the leading order term in a Pad\'{e} approximated $Q$ (reported in Eq.~(30) of Ref.~\cite{bazant2004diffuse}) 
for overlapping double layers ($\kappa L\ll1$) is $5L^2/(12D)$.
Interestingly, this prefactor $5/12=0.417$ for $Q$ is much closer to the correct value $4/\pi^2$ than the prefactor $1/3$ for $q$.

To put our findings for overlapping double layers into context, 
it is instructive to consider \eqr{eq:exactq} in the limit $\kappa L\to 0$,
which, filling in $\mathcal{M}_{j}=(2j-1)\pi/2$, simplifies to
\begin{align}\label{eq:Beunischargedenisty}
\frac{q(x,t)}{2\rhos\Phi}=&\frac{x}{L}
+\frac{8}{\pi^2}\sum_{j\ge1}
  \frac{\displaystyle{(-1)^{\displaystyle j}\sin\left[(2j-1)\frac{\pi x}{2L}\right]} } {(2j-1)^2}
\exp{\left[-\frac{t}{\tilde{\tau}_{j}}\right]},
\end{align}
with $\tilde{\tau}_{j}=4L^2/\left[D\left((2j-1)^2\pi^2\right)\right]$.
Equation~\eqref{eq:Beunischargedenisty} is equivalent to the ionic charge density 
that follows from Eq.~(51) of Ref.~\cite{beunis2008dynamics}. 
Moreover, in this limit $\kappa L\to 0$, \eqr{eq:exactEfield} predicts an unscreened electric field, $E(x)\simeq\Psi/L$,
which was precisely the assumption made in Ref.~\cite{beunis2008dynamics} to obtain their expression for the ionic densities.
We note that, given an unscreened electric field,  the term $D\kappa^2 q$ drops out of \eqr{eq:DebyeFalkenhagen}, 
leaving behind an ordinary diffusion equation for $q$. 
Therefore, timescale $4L^2/(D\pi^2)$ found for thick double layers also appears frequently 
as the timescale with which other diffusing systems relax; for neutral salt diffusion it has been known for over a century
\cite{rosebrugh1910mathematical,mckay1930diffusion}. 

We note that the late-time transients to the DF equation were also studied in 
Refs.~\cite{alexe2006transient,  barbero2006role, barbero2007electrical, beunis2007diffuse}. 
Our \eqr{eq:exactq} a posteriori justifies the ansatz made there
of a local ionic charge density whose 
position and time dependence are factored. With that ansatz, the relaxation times 
reported in those works follow from eigenvalue problems that have essentially the same form as our
\eqr{eq:realmpole} (for the parameters considered in this article). 
The higher order solutions $M_{j\ge2}$, important at short times, were mentioned but not elaborated on in Refs.~\cite{alexe2006transient, barbero2006role}. 
Even if all these modes would be determined, it is not obvious how to determine all the coefficients in the infinite sums in Eqs.~\eqref{eq:exactq}, \eqref{eq:exactcurrent}, and \eqref{eq:exactEfield}.

\subsection{Plots of the ionic charge density, ionic current density, and electric field}
\begin{widetext}

\begin{figure}[!h]
\centering
\includegraphics[width=\textwidth]{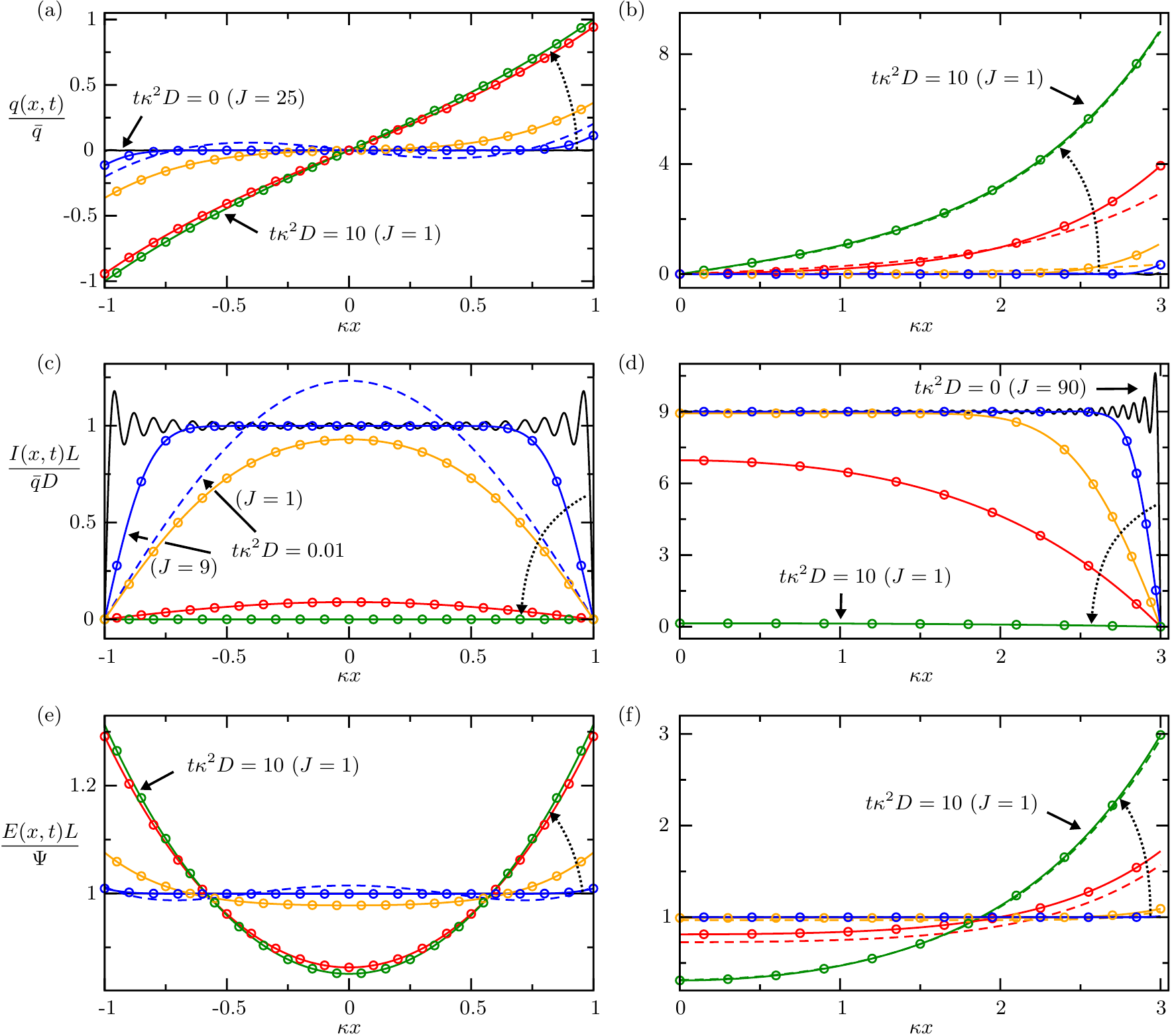}
\caption{The ionic charge density [(a) and (b)], ionic current density [(c) and (d)] and electric field [(e) and (f)] 
	as found via inverse Laplace transformations 
	[solid curves: Eqs.~\eqref{eq:exactq}, \eqref{eq:exactcurrent}, and \eqref{eq:exactEfield}, respectively] 
	and via numerical inverse Laplace transformation (open circles) \cite{hollenbeck1998invlap} for $n\equiv\kappa L=1$ [(a), (c), and (e)]
	and $\kappa L=3$ [(b), (d), and (f)], for which we only show the right half of the system.
	We evaluate these equations at $t\kappa^2 D=\{0, 0.01, 0.1, 1, 10\}$ 
	(black, blue, orange, red, green): the dotted arrows indicate the direction of increasing time. 
	For these successive times, we truncate the sums in Eqs.~\eqref{eq:exactq}, \eqref{eq:exactcurrent}, and \eqref{eq:exactEfield} after $J$ terms
	with $J=\{25, 5, 2, 1, 1\}$ [(a) and (e)], $J=\{25, 9, 4, 2, 1\}$ [(b), (c), and (f)], and $J=\{90, 18, 8, 2, 1\}$ (d).  
	The meaning of dashed lines differs among the subfigures: they indicate the respective quantities 
	at $t\kappa^2 D=0.01$ using $J=1$ [(a), (c), and (e)], and the Pad\'{e} approximation \eqr{eq:padeionicchargedensity} (b) 
	and Eq.~40 of Ref.~\cite{golovnev2011analytical} (f) at times $t\kappa^2 D=\{0, 0.01, 0.1, 1, 10\}$. 
      \label{fig:exactioniccharge}
    }
\end{figure}

\end{widetext}

Truncating the sums in Eqs.~\eqref{eq:exactq}, \eqref{eq:exactcurrent}, and \eqref{eq:exactEfield}
after a suitably chosen number $J$ of modes, in Fig.~\ref{fig:exactioniccharge} 
we plot (solid curves) the position dependence  of the ionic charge density, ionic current density, and electric field
 at several times, for two degrees of double layer overlap ($n\equiv\kappa L=1$ and $\kappa L=3$).
These two values correspond to either of the two cases of \eqr{eq:mathcalM} for which $\mathcal{M}_{1}=M_{1}$ ($n<\sqrt{3}$) or 
$\mathcal{M}_{1}=i\tilde{m}_{1}$ ($n>\sqrt{3}$). 
Also shown are data (circles) of numerical inverse Laplace transformations of 
Eqs.~\eqref{eq:localchargedensity}, \eqref{eq:currentxs}, and \eqref{eq:Exs} that were 
obtained by means of the 't Hoog algorithm \cite{de1982improved,hollenbeck1998invlap}.
The physical quantities presented in Fig.~\ref{fig:exactioniccharge} were 
nondimensionalized with different combinations of system parameters all involving the electrode potential $\Psi$. 
One should keep in mind that all those quantities were obtained within the Debye-Falkenhagen approximation, 
whose validity is restricted to the regime of small applied potentials $e\Psi/\kbt\ll1$.

At $t=0$, the exponents in the sums in Eqs.~\eqref{eq:exactq}, \eqref{eq:exactcurrent}, and \eqref{eq:exactEfield} are all unity.
However, for the ionic charge density and the electric field, whose coefficients become smaller with $j$, these sums can again be truncated at a finite number of terms: Their initial values should lie at $q(x,t=0)=0$ and $E(x,t=0)=\Psi/L$, respectively, 
which is decently approximated by the black lines ($J=25$).
Conversely, for the ionic current density, such a good behavior is not obtained.
Because  $\partial_{x}q(x,t=0)=0$, \eqr{eq:ioniccurrent} predicts an Ohmic response 
$I/(\bar{q}D)=n^2 E/\Psi$ at the moment of applying the potential,
explaining the relation between the plateau heights in the bulk as observed in Figs.~\ref{fig:exactioniccharge}(d) and \ref{fig:exactioniccharge}(f).
Simultaneously, $I(x=\pm L,t)=0$ must be satisfied. The combination of a nonzero constant ionic current density in the bulk ($x\neq\pm L$), 
and a vanishing ionic current density at the boundaries ($x=\pm L$) gives rise to the Gibbs phenomenon, where a discontinuous function approximated 
by a Fourier series overshoots the step height by $18\%$, which is indeed observed in Fig.~\ref{fig:exactioniccharge}(c) and \ref{fig:exactioniccharge}(d).
Hence, for $t=0$ the sum in \eqr{eq:exactcurrent} may not be cut at any \mbox{finite $J$}.

In Figs.~\ref{fig:exactioniccharge}(a), \ref{fig:exactioniccharge}(c) and \ref{fig:exactioniccharge}(e) 
we observe that, at small, nonzero times $t\kappa^2 D=0.01$, 
 the dashed lines ($J=1$) do not accurately reproduce
the data of the numerical inversions, whereas the solid lines ($J=5$ [Figs.~\ref{fig:exactioniccharge}(a) and \ref{fig:exactioniccharge}(e)]
and $J=9$ [Fig.~\ref{fig:exactioniccharge}(c)]) do. 
Importantly, here ``short'' does not imply timescales inaccessible to experiment;
for large $L$, the $j\ge2$ modes might decay sufficiently slow to be measurable experimentally.
On the other hand, the Debye-Falkenhagen equation (and its solution presented here) does not capture the fast relaxation processes associated 
with molecular vibrations and rotations, nor electron transfer processes that occur on very short time-scales.

The relaxation timescale $\tau_{j}$ become smaller with increasing $j$. Therefore, the modes with $j\ge2$ 
in the sums of Eqs.~\eqref{eq:exactq}, \eqref{eq:exactcurrent}, and \eqref{eq:exactEfield} all 
decay faster than the $j=1$ mode, and are important merely at small times.
At large time we obtain very good agreement between the numerical inversions and our expressions, even for $J\le2$.
At extremely long times, neutral salt diffusion (neglected in this article) 
in the bulk has been reported to affect the ionic current density: instead of exponentially decaying, 
the ionic current density then decays with a power law \cite{beunis2007power, marescaux2009impact}.

Next to the aforementioned exact and numerical results, in Fig.~\ref{fig:exactioniccharge}(b) we show the Pad\'{e} 
approximated ionic charge densities [\eqr{eq:padeionicchargedensity}] 
with dashed lines. These approximations describe the decay of the current fairly well, but they are not nearly as accurate as the expression for $q(x,t)$
derived here.
Moreover, Ref.~\cite{golovnev2011analytical} has also derived a solution for the electric field $E(x,t)$. 
However, we found no agreement between that expression [Eq.~(40) of Ref.~\cite{golovnev2011analytical} shown as dashed curves in Fig.~\ref{fig:exactioniccharge}(f)]
and our \eqr{eq:exactEfield} nor to the numerical Laplace inversion, except in the long and short time limits. The discrepancy can be traced back to the argument leading to Eq.~(22) in  Ref.~\cite{golovnev2011analytical}.

\section{Conclusion}\label{sec:conclusion}
We have presented expressions for the ionic charge density [\eqr{eq:exactq}], ionic current density [\eqr{eq:exactcurrent}], 
and electric field [\eqr{eq:exactEfield}] in a model electric double-layer capacitor (EDLC) in response to a small, suddenly applied potential. 
In particular, \eqr{eq:exactq} is the solution to the Debye-Falkenhagen equation,
which is easily solved in Laplace transformed (frequency) representation $\hat{q}(x,s)$, though leaving behind
a Laplace back transformation problem ($\mathcal{L}^{-1}\left\{\hat{q}\right\}$) that has been unsolved for over a decade.
So-called Pad\'{e} approximations to the Laplace-transformed $\hat{q}(x,s)$
can be readily inverted, but such approximate solutions to $q(x,t)$ have a number 
of shortcomings, including position-dependent decay rates. Moreover, by these methods, 
different decay rates are found among other, related EDLC observables.

In this article we have solved the problem $\mathcal{L}^{-1}\left\{\hat{q}\right\}$,
and moreover found exact expressions for the concomitant ionic current density and the electric field.
These solutions display none of the above-mentioned problems, and are 
in excellent agreement with numerical inverse Laplace transformations 
at all nonzero times and system sizes that we have studied.
Equations~\eqref{eq:exactq}, \eqref{eq:exactcurrent}, and \eqref{eq:exactEfield} are exact,
provided that we have identified all the poles of the functions to be inverted,
which we cannot prove at present, but which is supported by our systematic numerical investigation of the function $\hat{f}(m,n)$ 
[cf.~\eqr{eq:denominator}] in the plane of complex $m\in\mathbb{C}$.

Since, in fact, $\hat{f}(m,n)$  has an infinite number of poles, Eqs.~\eqref{eq:exactq}, \eqref{eq:exactcurrent}, and \eqref{eq:exactEfield} 
all contain infinite sums, 
whose coefficients depend on $\mathcal{M}_{j}$, the solutions to a transcedental equation [\eqr{eq:mpole}].
Moreover, each term of these sums decays exponentially with time, where, importantly,
the same timescales $\tau_{j}=L^2/[D\left(n^2+\mathcal{M}_{j}^{2}\right)]$ 
appear in all considered quantities. 
At nonzero times, one typically only needs the first few terms of these sums 
to highly accurately approximate the ionic charge density, ionic current density, and electric field.
The expression for the ionic charge density and electric field work even at the moment of applying the potential. 
The same is not true for the ionic current at $t=0$, where the Gibbs phenomenon occurs 
if the sum is truncated at any finite number of terms.

While we shortly discuss one extension of our model problem (including a finite Stern layer) in Appendix.~\ref{appendix:stern},
future work can extent on this article by considering, e.g., nonisothermal electrolytes, other time-dependent potentials (linear, sinusoidal, etc.), or adsorption or Faradaic reactions at the electrode surfaces. 
Exact results for those quantities can in turn be compared to a large body of published 
work.

\begin{appendix}

\section{Stern layer}\label{appendix:stern}
To describe Stern layers of thickness $\ls$, we extend our model setup such that the electrodes now lie at 
$x=\pm (L+\ls)$. The region $-L<x<L$ is still fully described by Eqs.~\eqref{eq:Poisson} to \eqref{eq:fixingA}, 
while within the Stern layers ($-L-\ls<x<L$ and $L<x<L+\ls$) the ionic charge density vanishes
and the potential is linear.
The potential at $x=-L$ amounts to $\Phi(x=-L, t)=\Phi(x=-L-\ls,t)+\ls \partial \Phi/\partial x|_{x=-L}$, 
where $\Phi(x=-L-\ls,t)$ is the step potential applied at $t=0$ onto the left electrode. With \eqr{eq:fixingA} we find 
\begin{align}\label{eq:Aconstantstern}
A_{1}&=\frac{\bar{q}m}{s}\left[\displaystyle{\frac{\sinh(m)}{m}+\left[\frac{m^2}{n^2}\left(1+\frac{\ls}{L}\right)-1\right]\cosh(m)}\right]^{-1},
\end{align}
which is equivalent to Eq.~(26) of Ref.~\cite{bazant2004diffuse}. 
Comparing \eqr{eq:Aconstantstern} to \eqr{eq:denominator}, we see that replacing $n$ by $\tilde{n}\equiv n/\sqrt{1+\ls/L}$ 
in Eqs.~\eqref{eq:exactq}, \eqref{eq:exactcurrent}, and \eqref{eq:exactEfield},
suffices to obtain expressions for the ionic charge density, ionic current density, 
and electric field in the case of nonvanishing Stern layers, where the the values of $\mathcal{M}_{j}(\tilde{n})$
 are now associated with the poles  of $\hat{f}(m,\tilde{n})$.
Note, however, that the explicit $n$ dependence
in the timescales $\tau_{j}$ [\eqr{eq:tauj}] remains unaltered as that dependence arises from the definitions of $n$ and $m$ themselves.

By the same arguments that led to \eqr{eq:tau1largen}, we find the long-time relaxation time for 
thin double layers $(n\gg1)$ and thin Stern layers $\ls\ll L$,
\begin{align}\label{eq:sterncorrection}
\tau_{1}&=\frac{L}{\kappa D(1+\ls \kappa)}\left[1+\mathcal{O}\left(\frac{1}{n}\right)\right].
\end{align}
in accordance with Eq.~(46) of Ref.~\cite{bazant2004diffuse} and Eq.~(5) of Ref.~\cite{beunis2007diffuse}. 
\end{appendix}

\end{document}